\documentstyle[aps,preprint,prb]{revtex}
\tightenlines
\begin{document}
\def\vk{\vec k} 
\def\br{{\bf r}}
\title{\bf Weak Localization Effect in Superconductors\\
 by Radiation Damage }
\author{Mi-Ae Park }
\address{Department of Physics,  University of Puerto Rico at Humacao,\\
 Humacao, PR 00791}
\author{Yong-Jihn Kim }
\address{Department of Physics,  Bilkent University,\\
 06533 Bilkent, Ankara, Turkey}
\maketitle
\begin{abstract}

Large reductions of the superconducting transition temperature $T_{c}$ and the 
accompanying loss of the thermal electrical resistivity
(electron-phonon interaction) due to radiation damage 
have been observed for several A15 compounds, Chevrel phase and Ternary 
superconductors, and $\rm{NbSe_{2}}$ in the high fluence regime.
We examine these behaviors based on the recent theory of weak localization 
effect in superconductors. 
We find a good fitting to the experimental data.
In particular, weak localization correction to the phonon-mediated 
interaction is derived from the density correlation function.
It is shown that weak localization has a strong influence on both the 
phonon-mediated interaction and the electron-phonon interaction, which 
leads to the universal correlation of $T_{c}$ and resistance ratio. 

\end{abstract}

\vspace{1pc}

\noindent

\vfill\eject
\section{\bf Introduction} 

Much attention has been paid to experimental and theoretical investigations
of the radiation effects on superconductors.$^{1-3}$
For the practical applications of the superconductors in the 
magnet coils of a fusion reactor,  the radiation response of the materials 
is important because they are subjected to irradiation.  
In theoretical side, the disorder effects in superconductors caused by 
irradiation are interesting.
The radiation effects in elemental type II superconductors showed that 
the superconducting transition temperature, $T_{c}$, does not change 
significantly for relatively high-fluence irradiations.$^{3}$ The slight
reduction was attributed to the reduction of the gap anisotropy. 
Subsequent annealing leads to the partial recovery of the $T_{c}$
changes, implying the importance of the microscopic details of the
disorder structures. We note that most elemental type II superconductors are 
radiation-tolerant. For instance, the He-4 dose which resulted in 
$\Delta \rho_{o} \sim 90\mu\Omega cm$ in $\rm{Nb_{3}Ge}$ causes   
$\Delta \rho_{o} \sim 2 \mu\Omega cm$ in Nb.$^{4}$    
Here $\rho_{o}$ denotes the residual resistivity.
On the other hand, A15 compounds show the universal large reductions of 
$T_{c}$ and the critical currents $I_{c}$ for high-fluence 
irradiations.$^{4-14}$ 
The residual resistivity $\rho_{o}$ also increases over $\sim 100\mu
\Omega cm$, indicating that A15 compounds are radiation-susceptible.
Note that the layered compound $\rm{NbSe_{2}}$, Ternary superconductors
$\rm{LuRh_{4}B_{4}}$ and $\rm{Er_Rh_{4}B_{4}}$,$^{15}$ and  Chevrel phase 
superconductors,$^{3}$ such as $\rm{PbMo_{6}S_{8}}$, $\rm{PbMo_{6}S_{7}}$, and 
$\rm{SnMo_{5}S_{8}}$ also show the large $T_{c}$ reduction in the high fluence 
regime.

In this paper we explain the microscopic mechanism underlying the  
universal large reductions of $T_{c}$ and $I_{c}$ in A15 compounds and other 
materials.  In particular, we stress the experimental observation of the 
correlation between the electrical-resistance ratio and $T_{c}$.$^{4,8,9,11}$
Testardi and his coworkers$^{4,8,9}$ considered Nb-Ge, $\rm{V_{3}Si}$, and 
$\rm {V_{3}Ge}$ for a variety of samples produced with differing
chemical composition, preparation conditions, and with varying amounts of
$^{4}\rm{He}$-induced defects. They found a close relation between
the resistance ratio and $T_{c}$ for those samples irrespective of the
manner how the disorder was achieved. Furthermore, they found that 
decreasing $T_{c}$ is accompanied by the decrease of the  thermal electrical 
resistivity (electron-phonon interaction).$^{4}$ 
It was also reported that tunneling 
experiments in $\rm{Nb_{3}Ge}$ and Nb-Sn clearly show a decrease of the 
electron-phonon coupling constant $\lambda$ accompanying the decrease of 
$T_{c}$ with disorder.$^{16-18}$. 

However, previous theoretical studies focused not on the changes of the 
electron-phonon interaction but on the smearing of the 
electronic density of states near the Fermi level $N(E_{F})$, the 
microscopic details of the disorder, the gap anisotropy, and the enhancement
of the Coulomb repulsion.$^{19-25}$ 
It is understandable that a consistent explanation of the existing 
experimental data was not possible in those prior theories. 
Recently, Kim and Overhauser$^{26}$ pointed 
out that Anderson's theorem$^{27}$ is valid only to the first power in the 
impurity 
concentration and the phonon-mediated interaction decreases exponentially
by Anderson localization, in agreement with the above
experimental findings.
As expected, it was shown that the same weak localization correction terms 
occur in both the conductivity and the phonon-mediated 
interaction.$^{28,29}$
Based on the reduced phonon-mediated interaction, we explain the universal
reductions of $T_{c}$ and $I_{c}$, and the universal correlation of $T_{c}$ and
resistance ratio in Sec. III.$^{30}$

Several comments are in order. 
(1) It is obvious that both impurity doping and 
irradiation (or implantation) can be used to study the disorder effects in 
superconductors and metals.
In particular, compensation of the $T_{c}$ reduction caused by magnetic 
impurities has been observed as a consequence of both radiation damage 
and ordinary impurity doping.$^{31-34}$ 
This compensation phenomenon has been predicted by Kim and 
Overhauser.$^{35,36}$
Recently, it has also been observed that impurity doping and/or 
ion-beam-induced damage in high Tc superconductors
cause the  metal-insulator transition and thereby suppress 
$T_{c}$.$^{37-40}$ These reductions may also be understood by the weak 
localization effect on superconductors. 
The only difference is the strong renormalization of the impurity potential
due to strong electron-electron interaction in high $T_{c}$ superconductors.
(2) Although Anderson's theorem suggests no change of the electron-phonon
interaction due to disorder how strong it is, there is overwhelming 
experimental evidence for the decrease of the electron-phonon interaction
in the strongly disordered samples and in the high fluence regime.
For instance, tunneling,$^{16-18}$ specific heat,$^{41}$ XPS,$^{42}$ 
loss of the thermal electrical resistivity,$^{4}$
and the correlation of $T_{c}$ and resistance ratio$^{4,8,9,11}$ manifest 
the decrease of the electron-phonon interaction when the electrons are weakly
localized. Weak localization leads to the decrease of the amplitude of the
electron wavefunction. As a result, the phonon-mediated matrix elements 
are also decreasing.$^{28,29}$ 
(3) Tunneling data do not show any enhancement of the Coulomb
repulsion.$^{16-18}$ In addition, the loss of the thermal electrical resistivity
with decreasing $T_{c}$ and the universal correlation between $T_{c}$ and
resistance ratio can not be explained in terms of the increase of the
Coulomb interaction.
(4) Irradiation also leads to the strong $T_{c}$ reductions in Chevrel phase
materials, such as $\rm{PbMo_{6}S_{8}, PbMo_{6}S_{7}}$, and
 $\rm{SnMo_{5}S_{8}}$,$^{3,43,44}$ and $\rm{NbSe_{2}}$,$^{43}$ at fluences
above $\sim 10^{18} n/cm^{2}$.
These materials are more radiation-sensitive than A15 compounds. 
It is clear that the origin of the strong $T_{c}$ reduction is not
related to the microscopic details of the disorder,  
but related to the universal nature of the electronic
state in the irradiated samples.

In Sec. II, we briefly review the experimental results of the radiation
damage effects on A15 compounds. The universal large 
reduction of $T_{c}$, the accompanying decrease of thermal electrical
resistivity, and the correlation of $T_{c}$ and resistance ratio will 
be emphasized. In Sec. III, weak localization correction 
on the phonon-mediated interaction is derived. The resulting $T_{c}$
decrease will be compared with experiments in Sec. IV.

\section{\bf Radiation Effects in A15 Compound Superconductors :\\ 
Universal $T_{c}$ Reduction  and Resistance Ratio}

Several A15 compounds have been investigated, including 
$\rm{Nb_{3}Sn}^{5-7,11-13,20}$, $\rm{Nb_{3}Al}^{6,7,10}$, 

\noindent
$\rm{Nb_{3}Ge}^{4,6-8,11,13}$, 
$\rm{Nb_{3}Ga}^{6,7}$, $\rm{Nb_{3}Pt}^{24}$,
$\rm{V_{3}Si}^{4,7,11}$, and $\rm{V_{3}Ge}^{4,11}$.
Both high-energy neutron$^{5,6,10}$ and other energetic charged particles,
such as protons,$^{14}$ $\alpha$ particles,$^{4,8,11,13}$ oxygens,$^{12}$ 
and electrons$^{20}$ were used to irradiate a variety of A15 compounds.
Table I summarizes the experimental results of the irradiation effects on 
A15 compound superconductors, Chevrel phases, and $\rm{NbSe_{2}}$.
Note that the large $T_{c}$ reductions are found in not only A15 compound
superconductors but also Chevrel phase superconductors, Ternary 
superconductors, and $\rm{NbSe_{2}}$, implying the universality of the phenomenon.

The  response of the superconducting properties of A15 compounds to irradiation
can be classified into the behavior at low fluences and at higher fluences. 
In the low-fluence regime, little or no change in $T_{c}$ occurs, while
universal large reductions of $T_{c}$ are observed for higher fluences. 
We focus on the universal $T_{c}$ reduction in this paper.
The boundary between the two regimes depends on the irradiating particles,
since the heavy ions give rise to more severe radiation damage.
For instance, the low-fluence regime corresponds to neutron fluence 
$< \ \sim 10^{18}n/cm^{2}$ and $^{4}He$ fluence 
$< \ \sim 10^{15}$ $^{4}He/cm^{2}$.
For much more higher fluences the saturation of $T_{c}$  is often found. 
It is noteworthy that the saturated $T_{c}$ state is accompanied by a saturated
value of the residual resistivity $\rho_{o}$.$^{4}$
Accordingly, the classification based on the residual resistivity
(not the fluence) may be more appropriate. In terms of the residual 
resistivity, the low-fluence regime corresponds to 
$\rho < \ \sim 10 \mu\Omega cm$ irrespective of the irradiating particles.
The saturations of $T_{c}$ and the residual resistivity are
easily understood in this classification scheme. 

From Table I, it is clear that the universal $T_{c}$ reduction is not crucially 
dependent on any specific irradiating particle,  any specific material and
any specific defect.
Many people noticed that the universal $T_{c}$ reduction is governed by the
total residual resistivity $\rho_{o}$ due to the radiation damage and the 
inherent damage present in the sample.$^{45, 4,8,9,11}$
Furthermore, the close relation between the $T_{c}$ decrease and resistance
ratio was established in $\rm{Nb-Ge}$, $\rm{V_{3}Si}$, and 
$\rm{V_{3}Ge}$.$^{4,8,9,11}$
The relation was also noticed in $\rm{Nb-O}$ solid solutions.$^{46}$
Testardi and his coworkers$^{4,8,9,11}$ reported that the correlation of 
$T_{c}$ and resistance ratio is independent of all sputtering conditions,
film thickness, composition, and radiation damage. 
This result implys that the defects produced during the irradiation are 
similar in their effect on $T_{c}$ to those produced during the film growth 
process. Consequently, the correlation of $T_{c}$ and resistance ratio is 
also universal. Until the resistance ratio is about 5, $T_{c}$ does not 
change much.  When it is smaller than 2, $T_{c}$ drops quickly. 
Finally superconductivity disappears if resistance ratio is around 1.  
Testardi et al.$^{4}$ also found that decreasing $T_{c}$ is accompanied by the loss of the thermal electrical resistivity (electron-phonon interaction),
which indicates the significant role of defects in both the superconducting 
and normal-state behavior. 
This finding, consistent with the correlation of $T_{c}$ and resistance
ratio, predicts the complete destruction of superconductivity for
resistance ratio less than 1 because of the complete loss of the 
electron-phonon interaction.

Other evidence for the decrease of the electron-phonon interaction in
the high fluence regime is the following: 
(1) By channeling measurements in single-crystal $\rm{V_{3}Si}$, 
Testardi et al.$^{47}$ also found that the radiation damage leads to the 
large increase of resistivity and the reduction of the electron-phonon 
interaction. (2) Viswanathan and Caton$^{48}$ reported the correlation of 
$T_{c}$ and residual resistivity in neutron-irradiated $\rm{V_{3}Si}$.  
(3) Tsuei, Molnar, and Coey$^{41}$ did a comparative study on the 
superconducting  and the normal-state properties of the amorphous and 
the crystalline phases of $\rm{Nb_{3}Ge}$. They found that the drastic 
reduction of $T_{c}$ 
is due to the changes in the strength of electron-phonon interaction.
Pollak, Tsuei, and Johnson$^{42}$ did an XPS study of the crystalline and
amorphous phases of $\rm{Nb_{3}Ge}$ and found that the crystalline phase has 
a higher $T_{c}$ because of the enhancement of the electron-phonon
coupling.

The response of the critical current $I_{c}$ also depends on fluence.
For low-fluence regime, $I_{c}$ decreases first with fluence, and then 
increases with increasing fluence, implying the importance of the 
flux-pinning mechanism.$^{49,50}$ 
For higher fluences, the universal reduction of $T_{c}$ drives down 
$I_{c}$.$^{5,51,52}$ The $I_{c}$ drop is field independent.$^{53}$ 

\section{\bf Weak Localization Correction to Phonon-mediated Interaction} 

In the presence of an impurity potential $U_{o}$, the Hamiltonian is 
given by
\begin{equation}
H=\int d{\bf r}\sum_{\alpha}\Psi_{\alpha}^{\dagger}({\bf r})[{p^{2}\over 2m}
+ U_{o}({\bf r})]\Psi_{\alpha}({\bf r}) -
V\int \Psi_{\uparrow}^{\dagger}({\bf r}) \Psi_{\downarrow}^{\dagger}({\bf r})
 \Psi_{\downarrow}({\bf r}) \Psi_{\uparrow}({\bf r})d{\bf r},
\end{equation}
where $\Psi^{\dagger}({\bf r})$ and  $\Psi({\bf r})$ are creation and 
annihilation operators for electrons.
In terms of the exact scattered states $\psi_{n}({\bf r})$, we expand the field
operator $\Psi_{\alpha}({\bf r})$ as
\begin{equation}
\Psi_{\alpha}({\bf r})=\sum_{n}\psi_{n}({\bf r})c_{n\alpha},
\end{equation}
where $c_{n\alpha}$ is a destruction operator of the electron.
Upon substituting Eq. (2) into Eq. (1), we find
\begin{equation}
H=\sum_{n}\epsilon_{n}c^{\dagger}_{n\alpha}c_{n\alpha} - \sum_{nn',n\not=n'}V_{nn'}
c_{n'\uparrow}^{\dagger}c^{\dagger}_{{\bar n'}\downarrow}c_{{\bar n}\downarrow}
c_{n\uparrow},
\end{equation}
where
\begin{equation}
V_{nn'}=V\int \psi_{n'}^{*}({\bf r})\psi_{{\bar n'}}^{*}({\bf r})
 \psi_{{\bar n}}({\bf r})\psi_{n}({\bf r})d{\bf r}.
\end{equation}
Here $\epsilon_{n}$ is the normal state eigenenergy and 
$\bar n$ denotes the time reversed partner of the scattered state $n$.
Eq. (4) was first obtained by Ma and Lee.$^{54}$

\subsection{Andersons' Theorem}

By a unitary transformation between the scattered states and the 
plane wave states, 
\begin{equation}
\psi_{n\alpha}=\sum_{\vk}\phi_{{\vk}\alpha}<{\vk}|n>,
\end{equation}
Eq. (4) can be rewritten as,$^{26}$ 
\begin{equation}
V_{nn'}=V \sum_{{\vk},{\vk'}{\vec q}}<-{\vk'}|n><{\vk}|n>^{*}
<{\vk}-{\vec q}|n'><-{\vec k'}-{\vec q}|n'>^{*}.
\end{equation}
Anderson$^{27}$ assumed that the transformed BCS part in Eq. (6) 
plays a much more
important role and each individual matrix element of the remaining 
interaction is so small as to be safely disregarded.
Then the normalization condition of the scattered states leads to
\begin{eqnarray}
V_{nn'}&\cong& V_{nn'}^{BCS}\nonumber \\
&=& V \sum_{{\vk},{\vk'}}|<{\vk}|n>|^{2}|<n'|\vk'>|^{2} \nonumber \\
&= & V,
\end{eqnarray}
which  is the essence of Anderson's theorem.

However, the remaining term,  
\begin{equation}
V^{non-BCS}_{nn'}=V \sum_{{\vk}\not=-{\vk'},{\vk'}{\vec q}}<-{\vk'}|n><{\vk}|n>^{*}
<{\vk}-{\vec q}|n'><-{\vec k'}-{\vec q}|n'>^{*},
\end{equation}
can not always be ignored.
As Anderson suggested, the above term is indeed negligible in 
low-fluence regime, where residual resistivity is smaller than 
$10\mu\Omega cm$. 
In fact, low-fluence regime corresponds to the dirty limit where 
$1/k_{F}\ell<0.1$. $k_{F}$ and $\ell$ denotes Fermi wave vector and mean free
path, respectively.
Whereas for higher fluences, the remaining term 
contribute significantly.   
In this regime, electron wave functions are weakly localized. 
Note that weak localization yields the well-known weak localization correction
to the conductivity.

Now we calculate $V_{nn'}$ including both the BCS and non-BCS terms.
To do that, we use Eq. (4) (not Eq. (6)) which is more transparent physically.
In the dirty limit, the exact eigenstates $\psi_{n}({\bf r})$ can be 
approximated by the incoherent superpositions of plane-wave states suggested 
by Thouless,$^{55}$ which leads to the Boltzmann conductivity.
The wavefunction, with energy $\hbar^{2}k_{n}^{2}/2m$, is written as
\begin{equation}
\psi_{n}({\bf r})=\sum_{\vk}a_{\vk}^{n}e^{i{\vk}\cdot{\bf r}}.
\end{equation}
The amplitudes $a_{\vk}^{n}$ are assumed to be independent normally-distributed random variables with the variance,
\begin{equation}
\overline{{a_{\vk}^{n}}^{*} a_{\vk'}^{n'}}\cong \delta_{nn'}\delta_{\vk\vk'}{\pi\over k_{n}^{2}\ell}{1\over (k-k_{n})^{2}+1/4\ell^{2}},
\end{equation}
for large $k_{n}\ell$.

Inserting Eq. (9) into Eq. (4) we obtain
\begin{eqnarray}
V_{nn'}&=& V\int |\psi_{n'}({\bf r})|^{2}|\psi_{n}({\bf r})|^{2}d{\bf r} \nonumber\\ 
&\cong& V\int|\sum_{\vk}a_{\vk}^{n}e^{i{\vk}\cdot{\bf r}}|^{2}
|\sum_{\vk'}a_{\vk'}^{n'}e^{i{\vk'}\cdot{\bf r}}|^{2}d{\bf r} \nonumber \\
&=&V\int \sum_{\vk}\overline{|a_{\vk}^{n}|^{2}}
\sum_{\vk'}\overline{|a_{\vk'}^{n'}|^{2}}d{\bf r}\nonumber \\
&=& V.
\end{eqnarray}
We have made use of Eq. (10) which eliminated the non-BCS term.
As a result, Anderson's theorem is proven for large $k_{n}\ell$ under the 
assumption.

\subsection{Weak Localization Correction}

For the high fluence regime, we may use the weakly localized wavefunctions
suggested by Kaveh and Mott.$^{56,57}$ 
For three dimensions, the weakly localized wavefunctions consist of power-law 
and extended wavefunctions,
\begin{equation}
\psi_{\vk}({\bf r})=Ae^{i{\vk}\cdot{\bf r}}+{B}{e^{ikr}\over r^{2}},
\end{equation}
where 
\begin{equation}
A^{2}=1-4\pi B^{2}({1\over \ell}-{1\over L}),\quad B^{2}={3\over 8\pi}{1\over k_{F}^{2}\ell}.
\end{equation}
$L$ denotes inelastic diffusion length.
We then write an eigenstate $\psi_{n}$ as
\begin{eqnarray}
\psi_{n}({\bf r})&=&\sum_{\vk}a_{\vk}^{n}\psi_{\vk}({\bf r})\nonumber \\
&=& \sum_{\vk}a_{\vk}^{n} (Ae^{i{\vk}\cdot{\bf r}}+{B}{e^{ikr}\over r^{2}}).
\end{eqnarray}
Comparing to the Thouless' wavefunction Eq. (9), Kaveh and Mott's
wavefunction includes the power-law component which originates from
the diffusive motion of the electrons.
Since the power-law wavefunction $1/r^{2}$ does not contribute to the current,
the conductivity is reduced as 
\begin{eqnarray}
\sigma^{3d}&=&\sigma_{B}A^{4}\nonumber \\
&=&\sigma_{B}[1-{3\over (k_{F}\ell)^{2}}(1-{\ell\over L})].
\end{eqnarray}
Similar situation occurs in the phonon-mediated interaction. The power-law
component does not contribute to the phonon-mediated matrix element either.
The reason is the following: since the power-law component peaks at some point,
its contribution to the {\sl bound} state of Cooper pairs far from the point is 
almost negligible. This is analogous to the insensitivity of the localized 
(bound) state with the change of the boundary conditions.$^{58}$
Accordingly, substitution of Eq. (14) into Eq. (4) leads to the weak localization correction to the phonon-mediated interaction, 
\begin{eqnarray}
V_{nn'}&=& V\int \sum_{\vk}\overline{|a_{\vk}^{n}|^{2}}
\sum_{\vk'}\overline{|a_{\vk'}^{n'}|^{2}}|\psi_{\vk}({\bf r})|^{2}
|\psi_{\vk'}({\bf r})|^{2}d{\bf r}\nonumber \\
&=& V\int |\psi_{\vk}({\bf r})|^{2} |\psi_{\vk'}({\bf r})|^{2}d{\bf r} \\
&\cong& VA^{4}\nonumber \\
&=&V [1-{3\over (k_{F}\ell)^{2}}(1-{\ell\over L})].
\end{eqnarray}
We have made use of the fact that Eq. (16) does not depend on $\vk$ or
$\vk'$.

We can also derive the weak localization correction term based on the
diffusive density correlation for the eigenstates. 
In the presence of impurities, the correlation function has a 
free-particle form for $t<\tau$ (scattering time) and a diffusive
form for $t>\tau$.$^{25}$ As a result, for $t>\tau$ (or $r>\ell$), 
one finds$^{59-62}$ 
\begin{eqnarray}
R &=& \int_{t>\tau} |\psi_{n}({\bf r})|^{2}|\psi_{m}({\bf r})|^{2}d{\bf r} \nonumber\\ 
&=& \sum_{\vec q}|<\psi_{n}|e^{i{\vec q}\cdot {\bf r}}|\psi_{m}>|^{2}\nonumber\\
&=&\sum_{1/L<\vec q<1/\ell}{1\over \pi\hbar N_{o}(E_{F})D{\vec q}^{2}}\\
&=&{3\over 2(k_{F}\ell)^{2}}(1-{\ell\over L}),
\end{eqnarray}
where $N_{o}(E_{F})$ and $D$ are the density of states and the diffusion
constant, respectively.
Then, the contribution from the free-particle-like density correlation 
is  
\begin{eqnarray}
V_{nm} &=& \int_{t<\tau} |\psi_{n}({\bf r})|^{2}|\psi_{m}({\bf r})|^{2}d{\bf r} \nonumber\\ 
&\cong& VA^{4}\nonumber \\
&=&V [1-{3\over (k_{F}\ell)^{2}}(1-{\ell\over L})],
\end{eqnarray}
with  $A^{2}=1-R$.$^{59}$
Since, only the free-particle-like density correlation contributes 
to the 
phonon-mediated interaction as explained above, we find the same weak 
localization correction to both the conductivity and the phonon-mediated
matrix element.  

The BCS $T_{c}$ equation is now,
\begin{equation}
T_{c}=1.13\omega_{D}e^{-1/\lambda_{eff}},
\end{equation}
where
\begin{equation}
\lambda_{eff}=N_{o}V [1-{3\over (k_{F}\ell)^{2}}(1-{\ell\over L})].
\end{equation}
The initial slope of $\Delta T_{c}$ is 
\begin{eqnarray}
\Delta T_{c} &\cong& {1\over \lambda}{3\over (k_{F}\ell)^{2}}(1-{\ell\over L}) \nonumber\\
&\propto& \rho_{o}^{2},
\end{eqnarray}
which is in good agreement with experiments.$^{13,63-65}$

\subsection{Strong-coupling Theory}

In the strong-coupling theory,$^{66,67}$ the electron-phonon coupling
constant is defined by
\begin{eqnarray}
\lambda&=&2\int{\alpha^{2}(\omega)F(\omega)\over \omega}d\omega \\
&=&N_{o}{<I^{2}>\over M<\omega^{2}>}.
\end{eqnarray}
Here $F(\omega)$ is the phonon density of states, and 
 $<I^{2}>$ and $<\omega^{2}>$ are the average over the Fermi surface of the 
square of the electronic matrix element and the phonon frequency, 
respectively.$^{67}$
For a homogeneous system with the Einstein model, it is written as 
\begin{eqnarray}
\lambda_{o}&=&N_{o}{I^{2}\over M\omega_{D}^{2}}\\
&=&N_{o}V \quad\quad\quad\quad  \rm{BCS \quad theory}
\end{eqnarray}

In the presence of impurities, weak localization leads to a correction to 
$\alpha^{2}$ or $<I^{2}>$, (disregarding the changes of $F(\omega)$ and $N_{o}$). 
From Eq. (25), one finds$^{68}$
\begin{equation}
\lambda=N_{o}{I^{2}\over M\omega_{D}^{2}}<\int
|\psi_{n}({\bf r})|^{2} |\psi_{m}({\bf r})|^{2}d{\bf r}>,
\end{equation}
which agrees with the BCS theory,
\begin{equation}
\lambda_{eff}=N_{o}V <\int|\psi_{n}({\bf r})|^{2} |\psi_{m}({\bf r})|^{2}d{\bf r}>.
\end{equation}

\subsection{Resistance Ratio}

According to Matthiessen's rule, resistivity $\rho(T)$ caused by static and 
thermal disorder is additive, i.e.,
\begin{equation}
\rho(T)=\rho_{o}+\rho_{ph}(T),
\end{equation}
where $\rho_{ph}$ is mostly due to electron-phonon scattering.
At high temperatures, the phonon limited electrical resistivity is given 
by$^{69}$
\begin{equation}
\rho_{ph}(T)={4\pi mk_{B}T \over ne^{2}\hbar}\int
{\alpha_{tr}^{2}F(\omega)\over \omega}d\omega,
\end{equation}
where $\alpha_{tr}$ includes an average of a geometrical factor
$1-cos\theta_{\vk\vk'}$.
Assuming $\alpha_{tr}^{2}\cong\alpha^{2}$, we obtain
\begin{eqnarray}
\rho_{ph}(T)&\cong&{2\pi mk_{B}T \over ne^{2}\hbar}\lambda_{eff}\nonumber\\
&\cong& {2\pi mk_{B}T \over ne^{2}\hbar} N_{o}{I^{2}\over M\omega_{D}^{2}}[1-{3\over (k_{F}\ell)^{2}}].  
\end{eqnarray} 
Note that decreasing $T_{c}$ is accompanied by the loss of the thermal 
resistivity $\rho_{ph}(T)$, in good agreement with experiment.$^{4}$   
The ternary superconductor $\rm{LuRh_{4}B_{4}}$,$^{15}$ also
shows the same behavior. 
The room temperature resistance ratio is then written as
\begin{eqnarray}
{\rho(300K)\over \rho_{o}}&=&{\rho_{o}+\rho_{ph}(300K)\over \rho_{o}}\nonumber\\
&\cong&1+{2\pi\tau \times 300K\over \hbar}\lambda_{eff}.
\end{eqnarray}
When $\lambda_{eff}$ goes to zero, the system is not superconducting and
resistance ratio becomes 1, which is in good agreement with experiments.$^{4,8,9,11, 46}$  
More details will be published elsewhere.

\section{\bf Comparison with experiments}

Wiesmann et al.$^{13}$ examined the $T_{c}$ change of  
vapor-deposited $\rm{Nb_{3}Ge}$ and $\rm{Nb_{3}Sn}$ as a function of 
$\alpha$ particle fluence. The 2.5-MeV $\alpha$ particles irradiated the 
samples, which were held at $30K$. The samples were then cooled and both $T_{c}$ 
and the residual resistivity $\rho_{o}$ were measured.
Figure 1 shows the dependence of $T_{c}$ on $\rho_{o}$ in $\rm{Nb_{3}Ge}$ and 
$\rm{Nb_{3}Sn}$.  
Thin lines are our theoretical results obtained from Eqs. (21) and
(22). We find good agreement between theory and experiment. 
The Debye temperature and $T_{co}$ (for the pure sample) are 
$\omega_{D}=302K$, $T_{co}=23K$ and 
$\omega_{D}=290K$, $T_{co}=18K$ for $\rm{Nb_{3}Ge}$ and $\rm{Nb_{3}Sn}$, 
respectively.
In the absence of experimental data for the inelastic diffusion length,
we used the same value of $L=\sqrt{D\tau_{i}}=\sqrt{\ell}\times 387\AA/T$ 
for both materials.$^{70}$
Here $\tau_{i}$ means the inelastic scattering time.
Since it is very difficult to evaluate $k_{F}\ell$ accurately,$^{71}$ we 
assumed that $\rho=100\mu\Omega cm$ corresponds to $k_{F}\ell=3.65$ and 
$3.60$ for $\rm{Nb_{3}Ge}$ and $\rm{Nb_{3}Sn}$ with the same value of
$k_{F}=0.3\AA^{-1}$.
These values also give a good fitting to the dependence of $T_{c}$
on the residual resistivity in impurity-doped samples.$^{72}$
This is persuasive evidence that the $T_{c}$ behavior is not crucially dependent
on any specific defect, rather its behavior is governed by the residual
resistivity.

Testardi and his coworkers$^{9}$ prepared about 130 Nb-Ge films and examined 
the dependence of $T_{c}$ on resistivity, resistance ratio, chemical 
composition, and sputtering conditions. 
The $\rm{Nb/Ge}$ ratios were in the range of $\sim 2.4 - 5.5$ and   
film thickness were about 2,000$\AA \sim$ 3,500$\AA$.
Only films which show the width of the superconducting transition less than
$\sim 2 - 3K$ were chosen to insure the macroscopic homogeneity of the
samples.  They found a universal correlation
of $T_{c}$ and resistance ratio irrespective of all sputtering conditions, 
composition, and specific nature of the disorder.$^{4,8,9,11,73}$
Figure 2 (a) presents a sampling of $T_{c}$-vs-resistance-ratio
data for 130 Nb-Ge films by them.
The correlation between $T_{c}$ and the resistance ratio is obvious.  
Resistance ratios less than 1 were generally found in films which are 
not superconducting, which agrees with theoretical expression, Eq. (33).
Our theoretical curve, which was obtained from 
Eqs. (21), (22), and (33), is also shown in the same figure.
We again find good agreement between the theoretical curve and
experiment. 
Since A15 compounds show deviations from Matthiessen's rule possibly 
due to saturation,$^{4,74}$ we adjusted $\rho_{ph}(T)$ to fit experimental 
values$^{4}$ at $300K$ in the following manner: 
$\rho_{ph}(300K)\cong 100\mu \Omega cm\times \lambda_{eff}(1-2.4/k_{F}\ell).$
We used the same values for $k_{F}, L$, and $\omega_{D}$ as in Fig. 1. 
But we found that $T_{co}=24K$ for 
the pure $\rm{Nb_{3}Ge}$ gives better fitting, which supports that sputtered 
films may have not yet achieved the highest possible $T_{c}$'s.$^{9}$ 

Poate et al.$^{8}$ irradiated superconducting Nb-Ge films by 2-MeV $\alpha$
particles and found a $T_{c}$-resistance correlation similar to that as-grown 
films.  Figure 2 (b) shows the correlations both for 
130 as-grown films$^{9}$ and for $\alpha$ particle irradiated films.$^{8}$
They lie nicely within the correlation band.
It indicates that the correlation of $T_{c}$ of resistance ratio is 
universal irrespective of how disorder is caused, e.g., by irradiation or 
substitutional alloying.
The correlation was also reported in $\rm{V_{3}Si}$ and $\rm{V_{3}Ge}$.$^{4}$

\section{\bf Discussion} 

It is clear that weak localization effect in superconductors caused by
impurity doping or radiation damage should be subjected to further 
experimental study. In particular, since the same weak localization 
correction term occurs both in the conductivity and the phonon-mediated 
interaction, comparative study of the normal and superconducting properties
of the samples will be beneficial.
It is noteworthy that Fiory and Hebard$^{65}$ found that both the 
conductivity and the transition temperature vary as $(k_{F}\ell)^{-2}$
for bulk amorphous $\rm{InO_{x}}$. 

The anti-localization effect of spin-orbit interaction will provide
more insights on weak localization effect in superconductors. In fact,
Miller et al.$^{75}$ found the compensation of $T_{c}$ decrease 
in highly disordered superconductors by adding impurities with
large spin-orbit scattering.
 
The loss of the thermal electrical resistivity $\rho_{ph}(T)$ 
(electron-phonon interaction) with decreasing $T_{c}$ needs more 
experimental study. In particular, we may consider samples satisfying the 
Matthiessen's 
rule, where the correlation of $T_{c}$ and resistance ratio is more
physically transparent.  We propose to investigate the usual low 
$T_{c}$ superconductors near the superconductor-insulator transition.$^{76}$
We expect to find the loss of the thermal electrical resistivity as
approaching the insulating regime.
Unfortunately, no systematic study is available yet.
Note that this behavior may provide a means of probing the phonon-mechanism
in exotic superconductors, such as, heavy fermion superconductors,
organic superconductors, and high $T_{c}$ cuprates. 

\section{\bf Conclusion} 

We have considered irradiation effects on A15 superconductors.
The universal large reductions of $T_{c}$ and $I_{c}$ due to radiation damage
has been explained by the weak localization of electrons.
Using weak localization correction to the phonon-mediated interaction
derived from the density correlation function, we calculated $T_{c}$
values which are in good agreement with experimental data.  
It is shown that weak localization decreases significantly
both the electron-phonon interaction and the 
phonon-mediated interaction, and thereby gives rise to 
  the universal correlation of $T_{c}$ and resistance ratio. 

\vspace{1pc}
\centerline{\bf ACKNOWLEDGMENTS}

YJK is grateful to Profs. Yun Kyu Bang and Bilal Tanatar for discussions and
encouragement. M. Park thanks the FOPI at the University of Puerto Rico-Humacao
for release time.

\vfill\eject

{\bf Table I.} Irradiation effects on A15 Compounds, Chevrel phases, and $\rm{NbSe_{2}}$ 

\vspace{2pc}

\begin{tabular}{lrrrrr} \hline \hline
{Sample} & \hspace{0.5pc} {Irradiating particle} &  {$T_{co}$} &\hspace{3pc}  {$\Delta T_{c}$} & \hspace{1pc} {Maximum fluence} & \hspace{1pc} {Reference}  \\ \hline

$\rm{Nb_{3}Ge}$  & $\alpha$ particle\hspace{2pc} & \hspace{1pc} $\sim  20K$ & $\sim 8K$ &  $10^{17} \alpha /cm^{2}$ & 8  \vspace{1pc}\\ 

 & neutron\hspace{2pc}  & \hspace{1pc} $\sim  20K$ & $ \sim 16K$ &  $5\times 10^{19} n/cm^{2}$ & 3  \vspace{1pc}  \\ 

$\rm{Nb_{3}Sn}$ & $\alpha$ particle\hspace{2pc} & \hspace{1pc} $\sim  18K$ & $\sim 15K$ &  $7\times 10^{17} \alpha/cm^{2}$ & 11  \vspace{1pc}\\ 

  & neutron \hspace{2pc} & \hspace{1pc} $\sim  18K$ & $\sim 7K$ &  $2\times 10^{19} n/cm^{2}$ & 5  \vspace{1pc}\\ 

  & electron \hspace{2pc} & \hspace{1pc} $\sim  17.8K$ & $\sim 3.8K$ &  $4\times 10^{20} el./cm^{2}$ & 20  \vspace{1pc}\\ 

$\rm{Nb_{3}Al}$  & neutron \hspace{2pc} & \hspace{1pc} $\sim  18K$ & $\sim 14K$ &  $5\times 10^{19} n/cm^{2}$ & 6  \vspace{1pc}\\ 

$\rm{Nb_{3}Pt}$  & neutron \hspace{2pc} & \hspace{1pc} $\sim  10.6K$ & $\sim 8.4K$ &  $3\times 10^{19} n/cm^{2}$ & 24  \vspace{1pc}\\ 

$\rm{V{_3}Si}$  & $\alpha$ particle\hspace{2pc} & \hspace{1pc} $ 16.8K$ & $\sim 14.5K$ &  $7\times 10^{17} \alpha /cm^{2}$ & 11  \vspace{1pc}\\ 

  &  neutron \hspace{2pc} & \hspace{1pc} $\sim 16.5K$ & $\sim 13.5K$ &  $2.5\times 10^{19} n/cm^{2}$ & 3  \vspace{1pc}\\ 
  
$\rm{V_{3}Ge}$ & $\alpha$ particle \hspace{2pc} & \hspace{1pc} $ 6.5K$ & $\sim 5.5K$ &  $5\times 10^{17} \alpha/cm^{2}$ & 11  \\ 
\hline
$\rm{PbMo_{6}S_{8}}$ & neutron \hspace{2pc} & \hspace{1pc} $12.8K$ & $\sim 8.6K$ &  $1\times 10^{19} n/cm^{2}$ & 3  \vspace{1pc}\\ 

$\rm{PbMo_{6}S_{7}}$ & neutron \hspace{2pc} & \hspace{1pc}  & $61\%$  &  $1.5\times 10^{19} n/cm^{2}$ & 3  \vspace{1pc}\\ 

$\rm{SnMo_{5}S_{8}}$ & neutron \hspace{2pc} & \hspace{1pc}  & $51\%$  &  $1.5\times 10^{19} n/cm^{2}$ & 3 \hspace{0.1pc} \\ 
\hline
$\rm{2H-NbSe_{2}}$ & neutron \hspace{2pc} & \hspace{1pc}  & $\sim 50\%$  &  $3\times 10^{18} n/cm^{2}$ & 3 \hspace{0.1pc} \\ \hline\hline

\end{tabular}

\begin{figure}
\caption{  Calculated $T_{c}$'s vs. residual
resistivity $\rho_{o}$ for $\rm{Nb_{3}Ge}$ and $\rm{Nb_{3}Sn}$. Experimental 
data are due to Wiesmann et al.,  Ref. 13.}
\end{figure}

\begin{figure}
\caption{ (a) Calculated $T_{c}$'s vs. resistance ratio for Nb-Ge.
The data points relate to about 130 films made with various sputtering voltage, deposition conditions, film thickness, crystal structure, and chemical 
composition. Data are from Testardi et al., Ref. 9.\\
 (b) $T_{c}$-resistance-ratio correlation band for 130 as-grown 
films with the values for damaged films superimposed. Data are from 
Poate et al., Ref. 8.}
\end{figure}


\begin{references}
\bibitem{1} S. V. Vonsovsky, Y. A. Izyumov, and E. Z. Kurmaev, {\sl Superconductivity of Transition Metals}, 
        Springer-Verlag, Berlin, (1982), ch. 9.
\bibitem{2} C. L. Snead, Jr and T. Luhman, {\sl Physics of Radiation Effects in Crystals}, eds. R. A. Johnson and A. N. Orlov, (Elsevier, 1986) ch. 6. 
\bibitem{3} A. R. Sweedler, C. L. Snead Jr, and D. E. Cox, in {\sl Treatise on Materials Science and Technology}, Vol. 14, eds Th. Luhman and D. Dew-Hughes
(Academic Press, New York, 1979) pp. 349-426.
\bibitem{4} L. R. Testardi, J. M. Poate, and H. J. Levinstein, Phys. Rev. B. {\bf 15}, 2570 (1977).  
\bibitem{5} R. Bett, Cryogenics,  {\bf 14}, 361 (1974).
\bibitem{6} A. R. Sweedler, D. S. Schweitzer, and G. W. Webb, Phys. Rev. Lett. {\bf 33}, 168 (1974) 
\bibitem{7}  A. R. Sweedler, D. E. Cox, and L. Newkirk, J. Electron.
Mater. ${\bf 4}$, \ \  883 (1975). 
\bibitem{8} J. M. Poate, L. R. Testardi, A. R. Storm, and W. M. Augustyniak, 
Phys. Rev. Lett. {\bf 35}, 1291 (1975).   
\bibitem{9} L. R. Testardi, R. L. Meek, J. M. Poate, W. A. Royer, A. R. Storm,
and J. H. Wernick, Phys. Rev. B {\bf 11}, 4304 (1975).
\bibitem{10} A. R. Sweedler and D. E. Cox, Phys. Rev. B {\bf 12}, 147 (1975). 
\bibitem{11} J. M. Poate, R. C. Dynes, L. R. Testardi, and R. H. Hammond, Phys. Rev. Lett. {\bf 37}, 1308 (1976).
\bibitem{12} G. Ischenko, H. Adrian, S. Kla\"umunzer, M. Lehmann, P. M\"uller,
H. Neum\"uller, and W. Szymczak, Phys. Rev. Lett. {\bf 39}, 43 (1977). 
\bibitem{13}   H. Wiesmann, M. Gurvitch, A. K. Ghosh, H. Lutz, K. W. Jones, A. N. Goland, and M. Strongin, J. Low. Temp. Phys. {\bf 30}, 513 (1978).
\bibitem{14} C. L. Snead, Jr., J. Nucl. Mater. {\bf 72}, 192 (1978). 
\bibitem{15} R. C. Dynes, J. M. Rowell, and P. H. Schmidt, in {\sl Ternary Superconductors}, ed. G. K. Shenoy, B. D. Dunlap, and F. Y. Fradin (North-Holland, Amsterdam, 1981), p. 169.  
\bibitem{16} K. E. Kihlstrom, D. Mael, and T. H. Geballe, Phys. Rev. B {\bf 29}, 150 (1984).
\bibitem{17} D. A. Rudman and M. R. Beasley, Phys. Rev. B {\bf 30}, 2590 
(1984). 
\bibitem{18} J. Geerk, H. Rietschel, and U. Schneider, Phys. Rev. B {\bf 30}, 459 (1984).  
\bibitem{19} R. C. Dynes and C. M. Varma, J. Phys. F {\bf 6}, L215 (1976).  
\bibitem{20} M. Gurvitch, A. K. Ghosh, C. L. Snead, Jr., and M. Strongin, Phys. Rev. Lett. {\bf 39}, 1102 (1977)  
\bibitem{21} H. Wiesmann, M.  Gurvitch, A. K. Ghosh, H. Lutz, O. F. Kammerer,
 and M. Strongin, Phys. Rev. B {\bf 17}, 122 (1978).
\bibitem{22} A. K. Ghosh, M. Gurvitch, H. Wiesmann, and M. Strongin, Phys. Rev. B {\bf 18}, 6116 (1978).
\bibitem{23} D. F. Farrell and B. S. Chandrasekar, Phys. Rev. Lett. {\bf 38}, 788 (1977). 
\bibitem{24} A. R. Sweedler, D. E. Cox, and S. Moehlecke, J. Nucl. Mater. {\bf 72}, 50 (1978). 
\bibitem{25} P. W. Anderson, K. A. Muttalib, and T. V. Ramakrishnan, Phys. Rev. B {\bf 28}, 117 (1983).  
\bibitem{26} Yong-Jihn Kim and A. W. Overhauser, Phys. Rev. B {\bf 47}, 8025 
     (1993).
\bibitem{27} P. W. Anderson, J. Phys. Chem. Solids {\bf 11}, 26 (1959).
\bibitem{28} Yong-Jihn Kim, Mod. Phys. Lett. B, {\bf 10}, 555 (1996).
\bibitem{29} Yong-Jihn Kim, Int. J.  Mod. Phys. B {\bf 11}, 1731 (1997).
\bibitem{30} Mi-Ae Park and Yong-Jihn Kim, Bull. Am. Phys. Soc. {\bf 44}, 1589 (1999). 
\bibitem{31} M. F. Merriam, S. H. Liu, and D. P. Seraphim, Phys. Rev. 
 {\bf 136}, A17 (1964).
\bibitem{32} G. Boato, G. Gallinaro, and C. Rizzuto, Phys. Rev. {\bf 148}, 
 353 (1966).
\bibitem{33} A. Hofmann, W. Bauriedl, and P. Ziemann, Z. Phys. B {\bf 46},
   117 (1982). 
\bibitem{34} M. Hitzfeld and G. Heim, Sol. Sta. Comm. {\bf 29}, 93 (1979). 
\bibitem{35} Yong-Jihn Kim and A. W. Overhauser, Phys. Rev. B {\bf 49}, 15779 
(1994).
\bibitem{36} Mi-Ae Park, M. H. Lee and Yong-Jihn Kim, Physica C {\bf 306}, 96 (1998).
\bibitem{37} J. M. Valles, Jr., A. E. White, K. T. Short, R. C. Dynes, J. P. Garno, A. F. Levi, M. Anzlowar, and K. Baldwin, Phys. Rev. B {\bf 39}, 11599 (1989). 
\bibitem{38} Y. Li, G. Xiong, and Z. Gan, Physics C {\bf 199}, 269 (1992). 
\bibitem{39} Y. Dalichaouch, M. S. Torikachvili, E. A. Early, B. W. Lee, C. L. Seaman, K. N. Yang, H. Zhou, and M. B. Maple, Solid State Commun. {\bf 65}, 1001 (1988). 
\bibitem{40} J. Fink, N. N\"ucker, H. Romberg, M. Alexander, M. B. Maple, J. J. Neumeier, and J. W. Allen, Phys. Rev. B {\bf 42}, 4823 (1990).
\bibitem{41} C. C. Tsuei, S. von. Molnar, and J. M. Coey, Phys. Rev. Lett. {\bf 41}, 664 (1978). 
\bibitem{42} R. A. Pollak, C. C. Tsuei, and R. W. Johnson, Sol. Sta. Com. {\bf 23}, 879 (1977). 
\bibitem{43} A. R. Sweedler et al., in {\sl Int. Conf. Radiat. Effects 
Tritium Technol. Fusion Reactors, Gatlinberg, Tennessee} CONF-750989, Vol II. p. 422 (1976).
\bibitem{44} B. S. Brown, J. W. Hafstrom, and T. E. Klippert, J. Appl. Phys. {\bf 48}, 1759 (1977).
\bibitem{45} A. K. Ghosh, H. Weismann, M. Gurvitch, H. Lutz, O. F. Kammerer, C. L. Snead, A. Goland, and M. Strongin, J. Nucl. Mater. {\bf 72}, 70 (1978).
\bibitem{46} C. C. Koch, J. O. Scarbrough, and D. M. Kroeger, Phys. Rev. B {\bf 9}, 888 (1974).  
\bibitem{47} L. R. Testardi, J. M. Poate, W. Weber, W. M. Augustyniak, and J. H. Barret, Phys. Rev. Lett. {\bf 39}, 716 (1977).
\bibitem{48} R. Viswanathan and R. Caton, Phys. Rev. B {\bf 18}, 15 (1978).
\bibitem{49} B. S. Brown, T. H. Blewitt, D. G. Wozniak, and M. Suenaga, J. Appl. Phys. {\bf 46}, 5163 (1975).
\bibitem{50} B. S. Brown, T. H. Blewitt, T. L. Scott, and D. G. Wozniak, J. Appl. Phys. {\bf 49}, 4144 (1978).
\bibitem{51} D. G. Schweitzer and D. M. Parkin, Appl. Phys. Lett. {\bf 24}, 333 (1974).
\bibitem{52} D. M. Parkin, and D. G. Schweitzer, Nucl. Technol. {\bf 22}, 108 (1974).
\bibitem{53} C. L. Snead Jr, and D. M. Parkin, Nucl. Technol. {\bf 29}, 264 (1976).
\bibitem{54} M. Ma and P. A. Lee, Phys. Rev. B {\bf 32}, 5658 (1985).
\bibitem{55} D. J. Thouless, Phil. Mag. {\bf 32}, 877 (1975).
\bibitem{56} M. Kaveh and N. F. Mott, J. Phys. C {\bf 14}, L177 (1981). 
\bibitem{57} N. F. Mott and M. Kaveh, Adv. Phys. {\bf 34}, 329 (1985). 
\bibitem{58} D. J. Thouless, Phys. Rep. {\bf 13 C}, 93 (1974). 
\bibitem{59} M. Kaveh, Phil. Mag. {\bf  51}, 453 (1985). 
\bibitem{60} W. L. McMillan, Phys. Rev. B {\bf 24}, 2739 (1981). 
\bibitem{61} B. L. Altshuler and A. G. Aronov, Zh. Eksp. Teor. Fiz. {\bf 77}, 2028 (1979) [Sov. Phys. JETP {\bf 50}, 968 (1979)]. 
\bibitem{62} E. Abrahams, P. W. Anderson, P. A. Lee, and T. V. Ramakrishnan,
Phys. Rev. B {\bf 24}, 6783 (1981). 
\bibitem{63} J. P. Orlando, E. J. McNiff, Jr., S. Foner, and M. R. Beasley, Phys. Rev. B {\bf 19}, 4545 (1979).  
\bibitem{64} S. J. Bending, M. R. Beasley, and C. C. Tsuei, Phys. Rev. B {\bf 30}, 6342 (1984). 
\bibitem{65} A. T. Fiory and A. F. Hebard, Phys. Rev. Lett. {\bf 52}, 2057 (1984). 
\bibitem{66} G. M. Eliashberg, Zh. Eksp. Teor. Fiz. {\bf 38}, 966 (1960) [Sov. Phys. JETP {\bf 11}, 696 (1960)]. 
\bibitem{67} W. L. Mc Millan, Phys. Rev. B {\bf 167}, 331 (1968). 
\bibitem{68} Yong-Jihn Kim, unpublished.  
\bibitem{69} G. Grimvall, {\sl The Electron-Phonon Interaction in Metals}, (North-Holland, Amsterdam, 1981) p.4. 
\bibitem{70} P. A. Lee and T. V. Ramakrishnan, Rev. Mod. Phys. {\bf 57}, 287 (1985).  
\bibitem{71} H. Gutfreund, M. Weger, and O. Entin-Wohlman, Phys. Rev. B {\bf 31}, 606 (1985).  
\bibitem{72} Yong-Jihn Kim and K. J. Chang, Mod. Phys. Lett. B{\bf 12}, 763 (1998).  
\bibitem{73} H. Lutz, H. Weismann, O. F. Kammerer, and M. Strongin, Phys. Rev. Lett. {\bf 36}, 1576 (1976).  
\bibitem{74} H. Wiesmann, M. Gurvitch, H. Lutz, A. Ghosh, B. Schwarz, M. Strongin, P. B. Allen, and J. W. Halley, Phys. Rev. Lett. {\bf 38}, 782 (1977).  
\bibitem{75} T. A. Miller, M. Kunchur, Y. Z. Zhang, P. Lindenfeld, and W. L. McLean, Phys. Rev. Lett. {\bf 61}, 2717 (1988).  
\bibitem{76} A. M. Goldman and Y. Liu, Physica D {\bf 83}, 613 (1995).  


\end{references}
\end{document}